# Wireless Sensor Networks Security: State of the Art

Mabrook Al-Rakhami[1], Student Member, IEEE, Saleh Almowuena[2], Student Member, IEEE

[1]College of Computer and Information Sciences, Research Chair of Pervasive and Mobile Computing, King Saud University, P.O. Box 51178, Riyadh 11543, Saudi Arabia
[2]1College of Computer and Information Sciences, Department of Computer Engineering, King Saud University, P.O. Box 51178, Riyadh 11543, Saudi Arabia

Corresponding author: Mabrook Al-Rakhami (e-mail: malrakhami@ksu.edu.sa).

**ABSTRACT** Wireless sensor networks (WSNs) have become one of the main research topics in computer science in recent years, primarily owing to the significant challenges imposed by these networks and their immense applicability. WSNs have been employed for a diverse group of monitoring applications, with emphasis on industrial control scenarios, traffic management, rescue operations, public safety, residential automation, weather forecasting, and several other fields. These networks constitute resource-constrained sensors for which security and energy efficiency are essential concerns. In this context, many research efforts have been focused on increasing the security levels and reducing the energy consumption in the network. This paper provides a state-of-the-art survey of recent works in this direction, proposing a new taxonomy for the security attacks and requirements of WSNs.

## I. INTRODUCTION

During the last decade, there was a significant technological advance in the areas of sensors and wireless communication, leading to the invention of wireless sensor networks (WSNs) [1]. This type of network can provide diverse functionalities such as monitoring, tracking, processing, and even controlling operations in the physical world. For instance, sensors can be interconnected to monitor and control environmental conditions in a forest or oceans, and they can be applied in healthcare systems or industrial machinery. Recent advances in microelectronics stimulated the development of small sensors, which were coupled to small devices provided with wireless communication, with low processing capacity and limited computing resources. The collection of these sensor nodes working cooperatively forms a WSN.

Despite the successful applications of WSNs in different fields, they have many drawbacks such as limitations in communication, processing, memory, and energy resources, which are, in most cases, due to their small batteries [2]. Fig. 1 shows the WSN forecast. Despite different predictions for the near future of WSNs (blue bars), all companies believe that the long-term trend market of WSNs will grow (red line). These predictions reveal the importance of WSNs, and we can conceive that our world will be significantly affected by the technologies involving WSNs [3].

Depending on the energy consumed by the sensor nodes, a WSN can operate for days or years; however, considering the difficulties of access to node locations, battery replacement may become impractical.

The security of the information processed by the nodes is of critical importance as these nodes can be deployed in diverse environments, such as for monitoring an oil well. Based on the nature of wireless communication, precedents are set for the most diverse types of information attacks. Thus, providing information security is a major problem in WSNs, which is the focus of several studies.

Cryptographic algorithms require significant energy consumption, processing, and memory. In general, cryptographic and authentication algorithms provide the services of confidentiality, integrity, and authentication, and hence, choosing the appropriate mechanisms is important for protecting the WSN data [4].

However, choosing an appropriate encryption algorithm would not entirely solve the energy consumption problem, as any implemented algorithm will influence the power consumption. Therefore, it is important to add security management to the system.

A security management system can act on a network by enabling and disabling security services and functions, whenever necessary, in response to events occurring over the network. This system can save network power if there is no indication or suspicion of intruders [5].

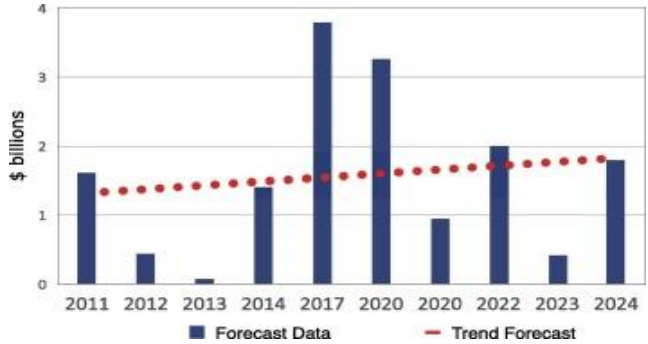

**FIGURE 1.** Wireless sensor network forecast

However, with the use of only an encryption algorithm and security level management, it is difficult to guarantee that the data is legitimate and did not suffer any type of attack. In this context, many authors [6-9] proposed models to estimate the level of security of the arriving data; consequently, the user can choose whether to use this data according to the level of security. Nevertheless, energy consumption can be improved by deploying an alternative secret sharing mechanism to the known public keys [10].

This paper provides an overview of the key challenges and trends in WSN security research, demonstrating solutions to problems and discussing the feasibility of using counterpoint solutions with increased power consumption, processing, and memory. Thus, this paper aims to provide an orientation of the techniques and mechanisms to be used in research related to WSN security.

## II. THEORETICAL BACKGROUND

### A. SECURITY OF WSNs
Security of WSNs should be considered in several applications. For example, if we place sensors in an oil well to detect drilling data in order to execute it in the best possible way, those responsible for collecting the data require the data to be reliable, with no interference, alteration, or inclusion of false data. Hence, a network should have the capacity to provide data integrity, confidentiality, authenticity, and availability, as well as be resistant to attacks [11]. This section presents the main security concepts of WSNs.

### B. BASIC CHARACTERISTICS
In this subsection, we will cover an overview of the essential characteristics required to ensure the security in WSNs, which include:

*Data Integrity*: Data integrity aims to ensure that all the original data attributes generated in the sensing node are maintained during the routing to the base station throughout the data life cycle [12].

*Confidentiality*: The right to access information must be provided only to the nodes with the authorization to access it [13].

*Authenticity*: It aims to ensure that the data remains the same as that produced by the source node, without any modification or mutation along the way [14].

*Availability*: Availability is fundamental, as this allows the data to be always available to authorized users [15].

### C. MOST COMMON ATTACKS ON WSN
Generally, all the attacks of WSNs are aimed at network malfunction or interruption of service. Therefore, the attacks are executed in several ways.

Table I presents a list of the common security threats and their solutions. In the following section, we briefly discuss the main attacks related to WSNs.

*1. Denial-of-service (DOS):*
This type of attack is used against web servers. It invalidates the network using an overload, aiming to consume all the memory and processing capacity of the network, in order to interrupt the services provided by the network [16-18]. DOS can be classified into the following attacks:

*Flood attack*: A network with unnecessary packets generates a large volume of traffic [19].

*Amplification attack*: A malicious node forges the victim's address and sends a large number of requests to the other nodes on behalf of the victim node; when all other nodes start responding to the requests, the victim node becomes congested [20].

*Exploiting protocol weaknesses*: This attack explores implementation failures in the victim's protocol. Accordingly, DoS attack easily interrupts the operation of the network, as the network devices are provided with low memory and processing capacity [21, 22].

*Wormhole attack*: It is a critical attack especially in the neighbor discovery phase, as this attack creates a tunnel between two nodes in different partitions of the network, leading the nodes to believe that they are neighbors while there are other knots between them, and accordingly, causing a convergence problem in the network [23, 24].

*Hello flood attack*: A fake node with high processing power and high signal power floods the network with HELLO messages, causing congestion throughout the network. Moreover, every other node believes that this false knot is a neighbor, and may also create false routes [25, 26].

*Black hole attack*: This is also called a sinkhole attack, where a malicious node shows false routes to the entire network, causing packets to pass through this false node before reaching the base station. This malicious node can discard or modify the routed packages [27, 28].

*2. Routing attacks:*
As the name of the attack suggests, the malicious node changes the routing, creating infinite loops or large deviations between nodes [29, 30].

*3. Jamming attack:*
A malicious node has a powerful transceiver configured to use the same frequency as the sensor nodes, which can occupy the communication channel with noise and prevent the sensor nodes from receiving messages [31-33].

TABLE I
WSN SECURITY THREATS AND SOLUTIONS [1]

| Security threats | Security requirement | Possible security solutions |
|---|---|---|
| Unauthenticated or unauthorized access | Key establishment and trust setup | Random key distribution, public key cryptography |
| Message disclosure | Confidentiality and privacy | Link/network layer, encryption, access control |
| Message modification | Integrity and authenticity | Keyed secure hash function, digital signature |
| Denial-of-service (DoS) | Availability | Intrusion detection, redundancy |
| Node capture and compromised node | Resilience to node compromise | Inconsistency detection and node revocation, tamper-proofing |
| Routing attacks | Secure routing | Secure routing protocols |
| Intrusion and high-level security attacks | Secure group management, intrusion detection, secure data aggregation | Secure group communication, intrusion detection |

*4. Sybil attacks:*
A malicious node assumes the pseudonymous identity of one or more legitimate nodes and subsequently executes various types of attacks on the network, including attacks on data aggregation, routing mechanisms, resource allocation, or distributed storage [23, 34-36].

*5. Message modification:*
A malicious node captures a message and retransmits it in an altered way [37].

*6. Data negligence and selective forwarding:*
The intruder node ignores the messages that must be sent or retransmitted [38].

*7. Node capture and compromised node:*
Most of the WSNs are operated in open environments. The small design of the sensors and the distributed nature of their deployment results in many vulnerabilities such as extracting cryptographic keys, tampering with the associated circuitry, changing the code of the sensors, and replacing sensors with malicious nodes that are controlled by the attacker[1].

**D. WSN SECURITY REQUIREMENTS**

*1. Cryptography:*
In general, cryptography involves applying a set of techniques, concepts, and methods to information in order to transform it into coded information, such that only the legitimate receiver of the information can decipher it [39].

*2. Encryption:*
It involves the process of transforming common information into coded information using a cryptographic algorithm. In the case of WSNs, data can be encrypted via an end-to-end or hop-to-hop process [40]. In the hop-to-hop process, encryption is performed each time the message passes through a different node until it reaches the base station. In this type of encryption, all neighbors must share the necessary keys for the process. In the end-to-end process, encryption is performed once per message; i.e., in a transmission, only the destination node and base station must encrypt/decrypt the message, rendering this process less expensive than the hop-to-hop process [41].

*3. Message signature:*
The signature of a message is intended to convince the recipient node that the message was generated by the sending node. The signature can be generated via the end-to-end processes and hop-to-hop processes [42].

*4. Key management:*
In the case of WSNs, public key sharing, which is commonly used in various types of networks, is unfeasible owing to the high cost of processing and power consumption. Sharing private keys is one way, but it renders the network very vulnerable because a network node can be hijacked, and the hijacker takes possession of the shared private key [43, 44].

*5. Intrusion detection system (IDS)*
An intrusion detection system aims to detect various types of malicious behaviors in the network, generating alerts from events. If an intrusion is detected, alerts are sent, and two types of response are possible: active and passive [45, 46]. In active response, malicious behavior is handled by the system itself, whereas in passive response, the system only generates reports such that the network administrator can observe and respond appropriately.

Owing to the collaborative and distributed nature of the WSNs, the ideal case is to use a collaborative and distributed

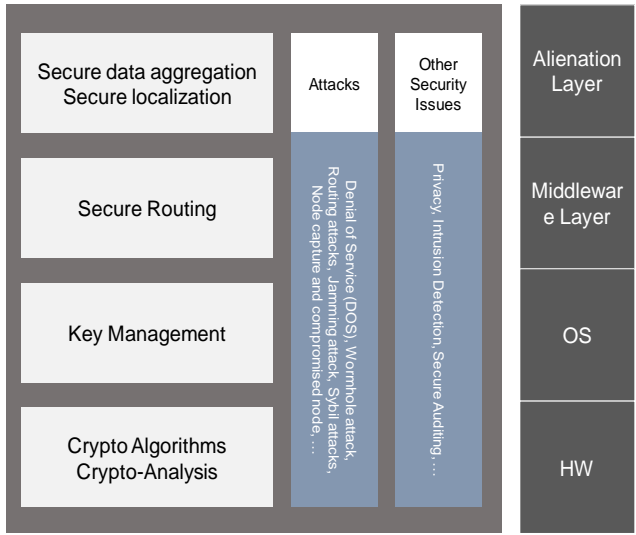

**FIGURE 2.** WSN security requirements

IDS. In these mechanisms, each sensor node monitors its neighbors for any suspicious behavior.

As soon as malicious activity is detected, neighbors exchange information about the suspect node. In this collaborative process, each neighbor sensor of a suspect node must indicate its point of view about this node to indicate if it is malicious or legitimate.

*6. Trust management system:*

Security management mechanisms exist to assess the reliability of sensor nodes, which can evaluate, maintain, and revoke trust between nodes. Usually, in the WSN context, the notion of trust is shown as follows: a node A relies on node B to execute process Y. This trust can be used for access control, secure routing, and intrusion detection [47, 48].

*7. Security and power management:*

Power management involves turning the device components on and off to maximize the life of the network via energy saving. As the use of security increases the energy consumption, security management becomes crucial [49]. This is to enable or disable security modules deployed in the network following the parameters defined by the network designer.

## III. RELATED WORK

In this section, we analyze studies related to the security of WSNs. The general taxonomy of recent works in this field for the last five years is divided based on a main category and sub-category as shown in Fig. 3. Table 2 summarizes all the references of our taxonomy.

### A. OVERVIEW

The authors in [215] addressed the global issues and challenges encountered in conserving power and security of WSNs. Power and security of WSNs are aggravated especially if the sensor is placed in environments where battery or sensor replacement is difficult. The authors discussed the main factors that increase energy consumption such as idle listening, collisions, overhearing, and overhead control packets. The authors explored energy conservation techniques currently being used such as duty cycling, mobility-based approaches, clustering, and game theory, and subsequently compared these techniques to demonstrate the approaches with the highest and lowest values of the following factors: idle listening, collisions, overhearing, control packets, energy saving, and time consumption. Eventually, they concluded that the priority-based technique was the best among the mentioned approaches.

TABLE 2
RECENT STUDIES OF WSN SECURITY [1]

| Categories | Sub-categories | Publications |
|---|---|---|
| Generic Survey | - | [50] [51] [52] [53] [54] [55] [56] [57] [58] [59] [60] [61] [62] [63] [64] [65] [66] [54] |
| Cryptography | Asymmetric | [67] [68] [69] [70] [4] [71] [72] [73] [74] [75] |
| | Symmetric | [76] [77] [78] [79] [80] [81] [82] [83] [84] [85] [86] [87] [88] [89] [90] |
| Secure Data Aggregation | - | [91] [92] [93] [94] [95] [96] [97] [98] [99] [100] [101] [102] [103] [104] [105] [106] [107] [108] [109] [110] [111] [112] |
| Secure Routing | - | [113] [114] [115] [116] [117] [118] [119] [120] [121] [122] [123] [124] [125] [126] [127] [128] |
| Secure Localization | - | [129] [130] [131] [132] [129] [133] [134] [135] [136] [137] [138] [139] [140] [141] [142] [129] |
| Key Management | - | [143] [144] [145] [146] [147, 148] [43] [149] [150] |
| Location Aware Security | - | [151] [152] [153] [154] [132] [155] [156] [157] [158, 159] |
| Attacks | Sybil | [160] [161] [162] [163] [164] [165] [166] [167] [168] [169] [170] [171] [172] [173] [174] [175] [176] |
| | Wormhole | [177] [178] [179] [180] [181] [182] [183] [184] [185] [186] [187] [188] [189] [190] |
| | DOS | [191] [192] [193] [194] [195] [196] [197] [198] [199] [200] [201] [192] |
| | Node Replication | [202] [203] [204] [205] [206] [207] [208] [209] [210] [211] |
| | Node Compromise | [212] [213] [102] [214] |

In [216, 217], the authors discussed WSNs attacks such as DOS, Sybil attack, wormhole attack, Hello flood attack, and black hole attack, proposed some security schemes for these attacks, and demonstrated their main features. Although the articles provide a direction for the mechanism to be used for energy conservation, they did not directly relate security to energy consumption. However, they described the main problems of energy conservation and security well, in addition to the difficulties in overcoming these problems, and proposed schemes for solving them.

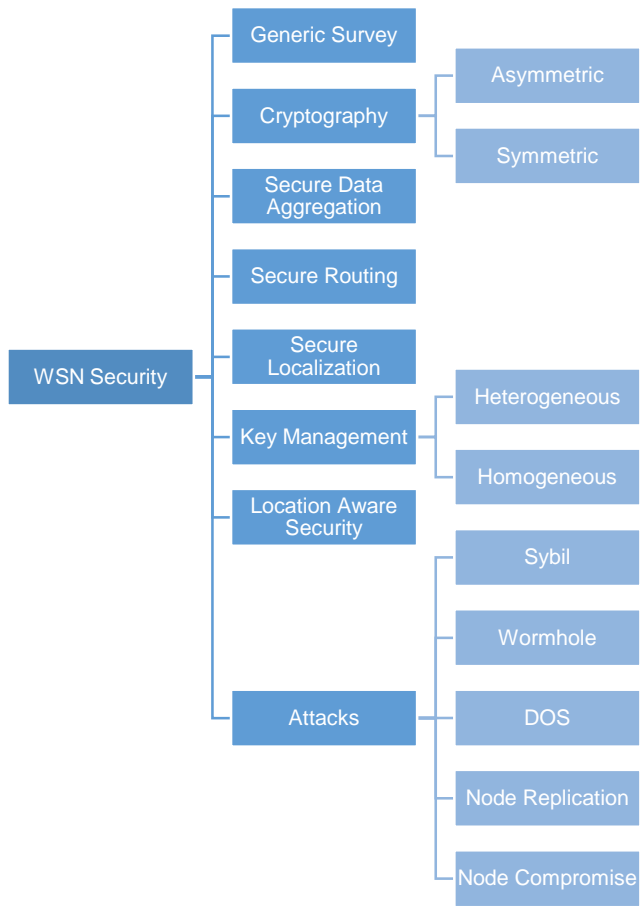

**FIGURE 3.** WSN security: taxonomy of recent works

## B. CRYPTOGRAPHY
The work in [218] demonstrated a performance evaluation of the security mechanisms for WSNs precisely, by analyzing the performance of the following cryptographic algorithms: AKGBE, novel stream cipher cryptosystem, fast and secure stream cipher, RC4, DES, and Blowfish. The authors presented many cryptography optimization algorithms for the key creation and encryption process, which is the ant colony optimization key generation-based image encryption method. The experiment was carried out using different parameters and the results were compared with the previous techniques to demonstrate the accuracy. The metrics considered in the evaluation of each algorithm were the maximum number of keys stored, battery capacity, and runtime.

In [219], the authored performed a performance evaluation of searchable symmetric encryption in WSNs. They examined four cryptographic hash functions—MD5, SHA1, SHA2, and SHA3—using the double hashing technique and truncating message digests. Moreover, they elected five symmetric encryption algorithms and two fast hash functions along with double hashing. They concluded that the best strategy for securing indexes was AES plus a fast FNV hash function and double hashing. The results showed that the integration of encryption algorithms with a fast hash function offers better results than using a cryptographic hash function alone. They concluded that using AES+FNV and the double hashing technique would improve the security level with acceptable power consumption.

## C. SECURE DATA AGGREGATION
As the sensor nodes can generate a large amount of redundant data, similar packets of multiple nodes can be aggregated to reduce the number of transmissions. Data aggregation is the combination of data from different sources according to some aggregation function. This technique has been used to achieve energy optimization and data transfer in a series of routing protocols. Signal processing methods can also be used for data aggregation. In this case, data aggregation refers to data fusion when a node is capable of producing a more accurate output signal by using some techniques to combine input signals and reduce noise in these signals. After the data collection and during its transfer to the main server, each sensor along the routing path cooperatively combines and secures the fragment messages.

Ensuring the security of in-network data aggregation has been the focus of many studies. The problem of data aggregation using encryption was introduced in [3] and further refined in [2]. The work in [94,95] proposed a secure information aggregation protocol to aggregate the data collected from the WSN nodes. The authors in [80] proposed symmetric-key-based homomorphic primitives for end-to-end secure data aggregation in WSNs to prevent redundant data transmission in data aggregation.

## D. KEY MANAGEMENT
Many WSN security implementations use encryption and key management schemes. The main problem is to make the management scheme adequately efficient such that, if an intruder captures or attacks a node, it is not possible for them to access all the keys of the network and thus the confidential information of the system. Different authors investigated and proposed solutions for the security of the WSNs by making assumptions regarding the capacity of the different nodes. As mentioned before, sensors have limited processing, storage, and power capabilities. However, it is possible to locate a small group of nodes whose resources are not limited, called a cluster leader. In these cases, the network is considered heterogeneous. In the case where all nodes have the same characteristics, the network is considered a homogeneous network.

## E. SECRET SHARING
Sharing mechanisms have been initiated by Shamir [220] to provide security for cryptographic keys. Thus, the work in [221] demonstrated a secret sharing-based key management scheme for sharing secrets in WSNs. It demonstrated the feasibility of collaborative security solutions on devices with good processing and storage capacity and contributed to the applications where collaboration among network nodes is required as a reliable mechanism.

The authors minimized the size of WSN clusters to adjust the overall energy consumption of the WSN. In contrast to some previous clustered architectural solutions, the notable improvement of this paper was that they discussed challenging security issues from the point of localizing key things based on the secret-sharing theory. The authors carried out the network key and cluster key and generated new keys from various polynomials using the Lagrange interpolation formula. Moreover, they invented a re-key mechanism in the cluster head election with low energy consumption.

## F. AUTONOMIC MANAGEMENT

Management involves handling something to obtain the best possible performance in several areas of performance, whereas autonomy indicates to do something independently where a concrete action (external intervention) is not required for making necessary changes. In this sense, the autonomic management in WSN is to save as much energy as possible, and accordingly, we can turn off the transistor of the device when it is idle as the radio is the component consuming the most energy [222].

Another way to improve energy consumption is by managing the security levels autonomously; by increasing or reducing the security using an automatic procedure when necessary, the consumption will be lower when the network feels safe, and security is not required. As sensor networks have low processing capacity and available energy, it is extremely important that the consumption of processing, memory, and energy is minimized, while not neglecting the security of the system. Therefore, this dilemma leads us to develop several ways to maintain a secure WSN.

One of the alternatives is by using a node management model in an autonomous way, where the nodes increase the level of security in the network (depending on the current state of the network). If there is no threat, the network can maintain itself with as little security as possible, and if there is an attack or threat, the level of security begins to grow at well-defined levels.

The work in [223] proposed a security management model for WSNs, including the selection of security components, description of management information, description of messages, and event definition in an autonomic network. Among the several security problems, the author considered the possibility of hop-to-hop and end-to-end encryption, the use of key management techniques, the existence of intrusion detection mechanisms, secure routing, aggregation of data, and node revocation scheme. Autonomic decisions were made by using an extension of their model, referred to as "MannaNMP," where the security components have dynamic configuration and messaging, and hence, they can be included, deleted, activated, and deactivated at run time using control messages.

These security components are based on intrusion detection events, and hence, the level of network security increases every time an attacker is detected. If the threat is perceived by the base station, it revokes the intruder node; if it is perceived by any of the other nodes, the base station increases the level of security. This level of security was divided by the author into four categories as follows:

- *Low*: No intruder detection at sensor nodes, no encryption, data fusion enabled;
- *Medium*: 10% of the nodes perform intrusion detection, updating authenticated end-to-end routes, hop-by-hop encryption enabled, data fusion enabled, alternate routes;
- *High*: 20% of nodes perform intrusion detection, end-to-end encryption enabled, authenticated hop-to-hop route updating, alternate routes, no data merging;
- *Critical*: 30% of nodes perform intrusion detection, no data fusion, end-to-end and hop-to-hop encryption enabled, authenticated hop-to-hop and end-to-end routes, alternate routes.

Notably, every time the level of security increases, there is a significant increase in the energy consumption and processing; therefore, if the network has a low amount of energy available, the security levels may be reduced even after the detection of intruders to ensure that the network does not stop functioning owing to power failure.

In order to validate the model described, the author performed simulations using a flat stationary network based on the model known as a hive for node arrangement, which varies between 50 and 1000. Each node has six known neighbors, all of which have the same computational resources and functionalities. Furthermore, the base station is defined. It is the source or destination of all the data and control packages. It has unlimited resources and cannot be violated.

After the evaluation of the security levels according to the criterion of energy consumption, the following average differences were obtained: from low to medium level, the increase in energy consumption was 9.9%; from medium to high, the increase was 18.7%; from high to critical, the increase was 8%; and from the low direct level to critical, the increase was 40%—thus confirming that the energy cost to use security is high, but via autonomic management, it can be improved. The evaluation emphasizes the biggest problem in WSNs, which is the energy efficiency and inclusion of security mechanisms. However, an interesting focus would be to evaluate the processing and consumption of RAM before the levels suggested by the author.

Despite extensive investigation of WSN security, it is not possible to guarantee a 100% secure network, owing to the unreliability of the communication channel, in addition to the exposure of the nodes to physical attacks. Several security mechanisms can be used to defend the network; however, if a security guarantee is required, the user is aware of the reliability of the network. Thus, the authors of [Sensor data security level estimation scheme for wireless sensor networks] presented the sensor data security level estimation scheme (SDSE), which is a new comprehensive scheme to estimate the level of data security of sensor networks based on their security mechanisms.

The level of security assists the user in deciding whether to use the data received. Moreover, the security level calculation can also be used to help experts make decisions about the evaluation of different security mechanisms and modification of the network settings to improve security (a sound and practical approach to quantifying security risk in enterprise networks + empirical analysis of system-level vulnerability metrics using actual attacks + an analytical hierarchical process-based risk assessment method for wireless networks).

According to the author, the SDSE model demonstrated in the article differs from the other related works in that the SDSE defines metrics that consider resilience and reliability based on the information revealing the current state of the network. The SDSE model uses stochastic methods as the metrics: the probability of cryptographic strength, probability of key management resilience, probability of legitimacy, and probability of delivery. The work considered four security mechanisms, two for prevention, i.e., cryptography and cryptographic key management, and two for detection, i.e., IDS and trust management system.

We can obtain the security level for data originating in a sensor node when it reaches the base station by the calculation of the metrics in each node by using the probabilities calculated. Each node on a given route has a degree of security, and the security level of the data will be the value of the lowest degree of security on the route. The author tested the prevention metrics using the known algorithm RC5 and observed that the greater the cryptographic strength of the algorithm, the more secure it is. Although with an increase in the time spent, the probability of cryptographic strength decreases, as more keys can be tested. In the case of resilience, it was observed that the higher the number of nodes captured, the lower the probability of resilience. Furthermore, the author observed that the probability of legitimacy decreases as more neighbors are required to detect the malicious node.

In general, the level of security is affected by all the parameters used in the metrics. The model proved to be feasible as the effective extraction of the parameters of the mechanisms was demonstrated with examples. Thus, the importance of the article is remarkable, but as the model was not effectively tested in WSNs with different security mechanisms, it is not possible to estimate the performance of the model.

## IV. CONCLUSION

In this work, a survey on the state-of-the-art works in WSNs was presented, and a new taxonomy for the security requirements was proposed. As future work, the scope of the work may be expanded by adding additional investigations and referring to more experiments carried out by authors to illustrate their conclusions with regard to WSNs security.

These tests can be added to the proposed security schemes, and a global analysis involving both problems and their possible solutions can be performed. The security mechanisms and performance metrics, such as network latency, should be evaluated. It is important to evaluate the impact of the different schemes and models in terms of energy consumption, processing, and memory in order to achieve viability to WSNs.

## ACKNOWLEDGMENT
This work was supported by College of Computer and Information Sciences, Department of Computer Science, King Saud University in the Kingdom of Saudi Arabia. We would like to express our thanks to Dr. Saleh Almowuena for his valuable notes and contribution in this work.